\documentclass[aps,prb,pdf,superscriptaddress,amsmath,amssymb,amsfonts,twocolumn,showpacs,nofootinbib]{revtex4-1}
\usepackage{amssymb}
\usepackage{amsmath}
\usepackage{graphicx}
\usepackage{dcolumn}
\usepackage{bm}
\usepackage{amsfonts}%
\usepackage{braket}
\usepackage{comment}

\usepackage[colorlinks=true,
linkcolor=blue,
filecolor=blue,      
urlcolor=blue,
citecolor=blue]{hyperref}
\providecommand{\U}[1]{\protect\rule{.1in}{.1in}}

\makeatletter
\newcommand*{\rom}
[1]{\expandafter\@slowromancap\romannumeral #1@}
\makeatother

\usepackage{soul}

\usepackage{xcolor}


\begin{document}

\title{Linear-response theory in Floquet systems}


\author{Ayan Pal}
\affiliation{Department of Physics, Division of Mathematical Physics, 
Lund University, Professorsgatan 1, 223 63, Lund, Sweden}
\affiliation{NanoLund, Lund University, Professorsgatan 1, 223 63, Lund, Sweden}

\author{Erik G. C. P. van Loon}
\affiliation{Department of Physics, Division of Mathematical Physics, 
Lund University, Professorsgatan 1, 223 63, Lund, Sweden}
\affiliation{NanoLund, Lund University, Professorsgatan 1, 223 63, Lund, Sweden}
\author{Ferdi Aryasetiawan}
\affiliation{Department of Physics, Division of Mathematical Physics, 
Lund University, Professorsgatan 1, 223 63, Lund, Sweden}
\affiliation{LINXS Institute of advanced Neutron and X-ray Science (LINXS),
Lund, Sweden}
\begin{abstract}
Nonequilibrium quantum physics greatly simplifies in the case of time-periodic Hamiltonians, since Floquet theory provides an analogue to Bloch's theorem in the time domain. Still, the formal properties of Floquet many-body theory remain underexplored. Here, we develop linear response theory for Floquet systems, in the sense that we have a time-periodic potential of arbitrary strength and a perturbatively small but non-periodic probing field. As an application, we derive the analogy of Fermi's Golden Rule and the photoemission spectrum of a many-electron system. As in the equilibrium case, the latter is related to the spectral function which is positive definite. We also analyze the parameter dependence of the controllable photoemission spectra by virtue of Floquet engineering.
\end{abstract}

\maketitle

\section{Introduction}

The dynamics of quantum many-particle systems
under time-dependent external fields
is one of the most active fields of research in chemistry and
condensed matter physics in the last few decades, both experimentally\cite{Broers2022Graphene, Broers2021Graphene}
and theoretically\cite{GRIFONI1998229,RevModPhys.86.779,Oka2009Hall_Effect}. A wide variety of new phenomena not accessible
in systems under equilibrium reveal themselves in the presence of
time-dependent fields. An intriguing example are photo-induced
phase transitions~\cite{fausti2011light,Bloch_2022}.
The possibility of controlling the phase of matter and 
changing the electronic~\cite{ECKSTEIN2021147108} as well as magnetic~\cite{PhysRevLett.76.4250, RevModPhys.82.2731, Trevisan2022PRL} properties of a material using a time-dependent external field is rather appealing, especially in the context of ultrafast (femtosecond) phenomena that access the intrinsic time-scales of the electrons in a solid.

A particular type of time-dependent external field of great interest is
a time-periodic one, which can be generated by, for example, lasers.
An interesting application
associated with time-periodic fields is Floquet design, in which quantum
systems are subjected to a chosen external drive to achieve certain
desired properties such as topological phases~\cite{rechtsman2013photonic,Trevisan2022PRL, Grushin2014CI,Yates2016CI,Li_2018CI}.
Despite the ubiquity of periodically driven quantum systems,
relatively little is known about the fundamental aspects of Floquet
theory, especially in the many-body context. 
Only very recently the positivity
of the spectral functions extracted from the retarded Floquet Green function
was rigorously established~\cite{Uhrig2019}.

In this article, a linear response theory for periodically
driven many-electron systems is developed in the wave function formalism,
which is an alternative to the Green function formalism
\cite{Stefanucci,Tsuji_2008}.
Here, the probing field is applied \emph{on top of} the periodic potential. Given a many-electron system in the presence of a time-periodic field, the theory provides a prescription on how to compute a first-order change in an observable when an additional \emph{arbitrary} probing field is applied. 

For a harmonic perturbation the corresponding Fermi golden rule for a Floquet system is derived. This is then applied to calculate the photoemission spectra and it is shown that the resulting
spectra can be related to a component of the Floquet Green function with positive definite spectra~\cite{Uhrig2019}. This provides a formal justification for identifying the density of states measured in photoemission experiments with the appropriate component of the Floquet Green function.
\begin{figure}[h!]
	\includegraphics[scale=0.22]{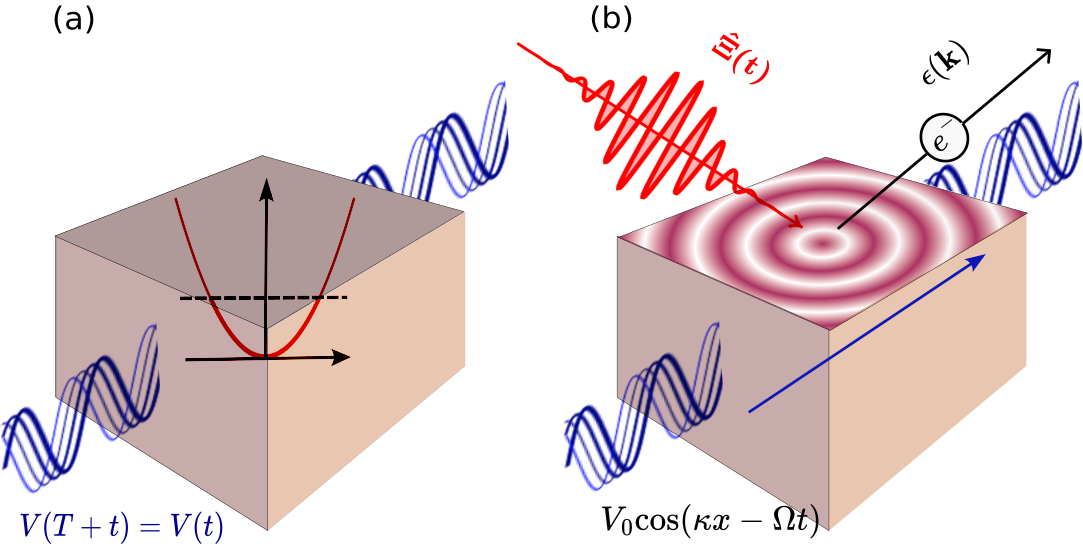}
	\caption{Schematic diagram of the formalism followed in this paper. (a) The system is initially prepared in a Floquet state with an external periodic driving field. (b) On top of that we apply a probing field that causes further electronic transitions or emission.} 
	\label{fig:Response Floquet}
\end{figure}

\section{Floquet theory}

A time-periodic Hamiltonian can be treated within Floquet theory, an analogue
of Bloch theory for periodic potential in space.
The one-particle time-dependent Schr\"{o}dinger equation is given by
\begin{equation}
\left[  \hat{H}_0(r,t)-i\partial_{t}\right]  \psi(r,t)=0, \label{TDSchr}%
\end{equation}
with time-periodicity, i.e., $\hat{H}_0(r,t+T)=\hat{H}_0(r,t)$. 
%
%
%
%
Here, $r$ denotes space and spin variables, $r=(\mathbf{r},\sigma)$.

According to Floquet's theorem, solutions to the time-dependent
Schr\"{o}dinger equation with a periodic potential in time have the form
\begin{equation}
\psi_{\alpha}(r,t)=e^{-i\varepsilon_{\alpha}t}\phi_{\alpha}(r,t),
\label{Floquet_soln}%
\end{equation}
where $\phi_{\alpha}(r,t)$ is periodic in time with period $T$,
\begin{equation}
\phi_{\alpha}(r,t+T)=\phi_{\alpha}(r,t),
\end{equation}
and it is an eigenfunction of the Floquet operator%
\begin{equation}
\hat{H}_{\text{F}}(r,t)=H_0(r,t)-i\partial_{t}%
\end{equation}
with eigenvalue $\varepsilon_{\alpha}$:
\begin{equation}
\hat{H}_{\text{F}}(r,t)\phi_{\alpha}(r,t)=\varepsilon_{\alpha}\phi_{\alpha
}(r,t). \label{Floquet_mode}%
\end{equation}
Here, $\psi_{\alpha}(r,t)$ and $\phi_{\alpha}(r,t)$ are called a
Floquet state and a Floquet mode, respectively. 
Eq. (\ref{Floquet_soln}) implies that
\begin{equation}
\psi_{\alpha}(r,t+T)=e^{-i\varepsilon_{\alpha}T}\psi_{\alpha}(r,t),
\end{equation}
which is the analogue of the Bloch theorem for solutions to the Hamiltonian of
a periodic crystal and $\varepsilon_{\alpha}$ is the analogue of the crystal
momentum whereas $T$ plays the role of the primitive translation vector in
space. From the group theory point of view, 
$\exp(-i\varepsilon_{\alpha}T)$ can be
regarded as a one-dimensional irreducible representation of the element $T$ of
the Abelian time translation group. 

It is clear that if $\phi_{\alpha}(r,t)$ is a Floquet mode then
\begin{equation}
\phi_{n\alpha}(r,t)=\phi_{\alpha}(r,t)e^{in\Omega t},
\end{equation}
where $\Omega=2\pi/T$, is also a Floquet mode with eigenvalue
\begin{equation}
\varepsilon_{n\alpha}=\varepsilon_{\alpha}+n\Omega. \label{eigval}%
\end{equation}
The corresponding Floquet state $\psi_{n\alpha}(r,t)$ is identical to
$\psi_{\alpha}(r,t)$:
\begin{equation}
\psi_{n\alpha}(r,t)=e^{-i\varepsilon_{n\alpha}t}\phi_{n\alpha}%
(r,t)=e^{-i\varepsilon_{\alpha}t}\phi_{\alpha}(r,t)=\psi_{\alpha}(r,t).
\end{equation}
The property in Equation (\ref{eigval}) allows us to restrict
$\varepsilon_{\alpha}$ as follows:
\begin{equation}
-\frac{\Omega}{2}\leq\varepsilon_{\alpha}<\frac{\Omega}{2}.
\end{equation}
The segment $[-\frac{\Omega}{2},\frac{\Omega}{2}]$ is the analogue of the
first Brillouin zone in a periodic crystal.
We have considered a one-particle Floquet system,
but the same properties are also valid in a many-particle Floquet
system.

\section{Floquet picture}

Let a time-periodic Hamiltonian $\hat{H}_0(t)$ be augmented by an arbitrary
external time-dependent perturbing field $\hat{\Xi}(t)$, which is not
in general periodic, i.e.,
\begin{align}
    \hat{H}(t)=\hat{H}_0(t) + \hat{\Xi}(t).
\end{align}
Define states in the Floquet picture (interaction picture) as follows:
\begin{align}
   |\Psi_\mathrm{F}(t)\rangle &= \hat{U}_0(t_0,t)  |\Psi(t)\rangle,
   \\
      \hat{U}_0(t,t_0) |\Psi_\mathrm{F}(t)\rangle &=  |\Psi(t)\rangle,
\label{eq:FloquetPic}
\end{align}
where
\begin{align}
    i\partial_t |\Psi(t)\rangle = \hat{H}(t) |\Psi(t)\rangle,
\label{eq:SchrEq}
\end{align}
and $\hat{U}_0(t,t_0)$ is the time-evolution operator associated with $\hat{H}_0(t)$,
which fulfills the equation of motion
\begin{align}
    i\partial_t \hat{U}_0(t,t_0) = \hat{H}_0(t) \hat{U}_0(t,t_0).
\label{eq:U0}
\end{align}
The time-evolution operator of the Floquet system, $\hat{U}_0(t,t_0)$,
can be expressed in terms of the Floquet modes as follows
\cite{GRIFONI1998229,Uhrig2019}:
\begin{align}\label{eq:U0}
    \hat{U}_0(t,t_0) = \sum_\alpha |\Phi_\alpha(t)\rangle 
    e^{-iE_\alpha(t-t_0)} \langle \Phi_\alpha(t_0)|,
\end{align}
The many-body Floquet modes $\Phi_{\alpha}(t)$ are the eigenvectors of the Floquet Hamiltonian
\begin{align}
    [\hat{H}_0(t)-i\partial_t] |\Phi_\alpha(t)\rangle
    = E_\alpha|\Phi_\alpha(t)\rangle.
\end{align}
$E_{\alpha}$ are the eigenvalues of the many-body Floquet Hamiltonian or the many-body quasi-energy. Operators in the Floquet picture are defined as
\begin{align}
    \hat{O}_\mathrm{F}(t)= \hat{U}_0(t_0,t) \hat{O}(t) \hat{U}_0(t,t_0),
\end{align}
where $\hat{O}(t)$ is an operator in the Schr\"odinger picture.

Taking the time derivative of Eq. (\ref{eq:FloquetPic}) yields
\begin{align}
    (i\partial_t \hat{U}_0(t,t_0))|\Psi_\mathrm{F}(t)\rangle
    +\hat{U}_0(t,t_0) i\partial_t |\Psi_\mathrm{F}(t)\rangle
    = i\partial_t |\Psi(t)\rangle.
\end{align}
Using Eqs. (\ref{eq:U0}) and (\ref{eq:SchrEq}) leads to
\begin{align}
    \hat{U}_0(t,t_0) i\partial_t |\Psi_\mathrm{F}(t)\rangle
    = \hat{\Xi}(t) \hat{U}_0(t,t_0) |\Psi_\mathrm{F}(t)\rangle.
\end{align}
Applying $\hat{U}_0(t_0,t)$ on both sides of the above equation gives
\begin{align}
    i\partial_t |\Psi_\mathrm{F}(t)\rangle 
    =\hat{\Xi}_\mathrm{F}(t)|\Psi_\mathrm{F}(t)\rangle,
\label{eq:SchrFloquet}
\end{align}
where
\begin{align}
    \hat{\Xi}_\mathrm{F}(t) = \hat{U}_0(t_0,t) \hat{\Xi}(t) \hat{U}_0(t,t_0).
\end{align}
Define the time-evolution operator in the Floquet picture as follows:
\begin{align}
    |\Psi_\mathrm{F}(t)\rangle 
    = \hat{U}_\mathrm{F}(t,t_0) |\Psi_\mathrm{F}(t_0)\rangle. 
\end{align}
From Eq. (\ref{eq:SchrFloquet}) one obtains
\begin{align}
    i\partial_t \hat{U}_\mathrm{F}(t,t_0) = \hat{\Xi}_\mathrm{F}(t)
    \hat{U}_\mathrm{F}(t,t_0),
\end{align}
with the formal solution
\begin{align}
    \hat{U}_\mathrm{F}(t,t_0) 
    = \mathcal{T} \exp{[-i\int_{t_0}^t dt_1 \hat{\Xi}_\mathrm{F}(t_1)] }.
\end{align}
$\mathcal{T}$ is the time-ordering. This provides a means for developing a perturbation expansion in the
perturbing potential $\hat{\Xi}(t)$ starting from a Floquet system. Although we have focused on a time-periodic $\hat{H}_0(t)$,
the derivation makes no assumption of a time-periodic Hamiltonian.

\subsection{Linear-response theory}

The expectation value of an observable associated with an operator
$\hat{O}$ as a function of time is given by
\begin{align}
    O(t)=\langle \Psi(t)| \hat{O}|\Psi(t)\rangle
    =\langle \Psi_\mathrm{F}(t)| \hat{O}_\mathrm{F}(t)|\Psi_\mathrm{F}(t)\rangle. 
\end{align}
To first-order in the perturbing field
\begin{align}
    |\Psi_\mathrm{F}(t)\rangle 
    &= \hat{U}_\mathrm{F}(t,t_0) |\Psi_\mathrm{F}(t_0)\rangle
    \nonumber\\
    &=\left[ 1 -i\int_{t_0}^t dt_1 \hat{\Xi}_\mathrm{F}(t_1) \right] 
    |\Psi_\mathrm{F}(t_0)\rangle
\end{align}
yielding
\begin{align}
    O(t)=& \langle \Psi_\mathrm{F}(t_0)| \hat{O}_\mathrm{F}(t)
    | \Psi_\mathrm{F}(t_0) \rangle
    \nonumber\\
    &-i\int_{t_0}^t dt_1 \langle \Psi_\mathrm{F}(t_0)| 
    [\hat{O}_\mathrm{F}(t), \hat{\Xi}_\mathrm{F}(t_1)] | \Psi_\mathrm{F}(t_0) \rangle.
\end{align}
For an operator that does not depend explicitly on time and assuming that
$| \Psi_\mathrm{F}(t_0) \rangle$ is a Floquet mode
$| \Phi_\alpha(t_0) \rangle$,
%
%
\begin{align}
   \langle \Psi_\mathrm{F}(t_0)| \hat{O}_\mathrm{F}(t)
    | \Psi_\mathrm{F}(t_0) \rangle
    &=\langle \Phi_\alpha(t_{0})| \hat{O}_\mathrm{F}(t)
    | \Phi_\alpha(t_{0}) \rangle
    \nonumber\\
    &=\langle \Phi_\alpha(t)|\hat{O}
    | \Phi_\alpha(t) \rangle.
\end{align}
Consider
\begin{align}
   & \langle \Phi_\alpha(t_0)| \hat{O}_\mathrm{F}(t)
   \hat{\Xi}_\mathrm{F}(t_1)    | \Phi_\alpha(t_0) \rangle 
    \nonumber\\
    &=\langle \Phi_\alpha(t_0)|\hat{U}_0(t_0,t) \hat{O} \hat{U}_0(t,t_1)
    \hat{\Xi}(t_1)\hat{U}_0(t_1,t_0)
    | \Phi_\alpha(t_0) \rangle 
    \nonumber\\
    &=\langle \Phi_\alpha(t)| \hat{O} \hat{U}_0(t,t_1)
    \hat{\Xi}(t_1) | \Phi_\alpha(t_1) \rangle e^{-iE_\alpha(t_1-t)}
    \nonumber\\
    &=\sum_\beta \langle \Phi_\alpha(t)| \hat{O} |\Phi_\beta(t)\rangle
    \langle\Phi_\beta(t_1)|
    \hat{\Xi}(t_1) | \Phi_\alpha(t_1) \rangle 
    \nonumber\\
    &\qquad\qquad \times e^{-i(E_\beta-E_\alpha)(t-t_1)},
\end{align}
where Eq. (\ref{eq:U0}) has been used to obtain the last line.
For a perturbing field coupled to the density,
\begin{align}
    \hat{\Xi}(t)=\int dr \varphi(rt)\hat{\rho}(r),
\end{align}
one obtains for the first-order change in the expectation value
\begin{align}
    \Delta O(t)&=-i\int dr' \int_{t_0}^t dt_1 
    \sum_\beta \langle \Phi_\alpha(t)| \hat{O} |\Phi_\beta(t)\rangle
    \nonumber\\
    &\times\langle\Phi_\beta(t_1)|
    \hat{\rho}(r') | \Phi_\alpha(t_1) \rangle 
    e^{-i(E_\beta-E_\alpha)(t-t_1)}\varphi(r',t_1)
    \nonumber\\
    &+i\int dr' \int_{t_0}^t dt_1 
    \sum_\beta \langle \Phi_\alpha(t_1)| \hat{\rho}(r') |\Phi_\beta(t_1)\rangle
    \nonumber\\
    &\times\langle\Phi_\beta(t)|
    \hat{O} | \Phi_\alpha(t) \rangle 
    e^{i(E_\beta-E_\alpha)(t-t_1)}\varphi(r',t_1).
\end{align}
The retarded linear response can be readily read off from the above expression.
Choosing $\hat{O}=\hat{\rho}(r)$ gives the retarded
linear density response:
\begin{align}
    R^\mathrm{R}(rt,r't')&=-i\theta(t-t')
    \sum_\beta \langle \Phi_\alpha(t)| \hat{\rho}(r) |\Phi_\beta(t)\rangle
    \nonumber\\
    &\qquad\times\langle\Phi_\beta(t')|
    \hat{\rho}(r') | \Phi_\alpha(t') \rangle 
    e^{-i(E_\beta-E_\alpha)(t-t')}
    \nonumber\\
    &+i\theta(t-t')
    \sum_\beta \langle \Phi_\alpha(t')| \hat{\rho}(r') |\Phi_\beta(t')\rangle
    \nonumber\\
    &\qquad\times\langle\Phi_\beta(t)|
    \hat{\rho}(r) | \Phi_\alpha(t) \rangle 
    e^{i(E_\beta-E_\alpha)(t-t')}.
\end{align}
This result agrees with a direct evaluation of the retarded
density-density correlation function:
\begin{align}
    R^\mathrm{R}(rt,r't')&=-i\theta(t-t')\langle \Phi_\alpha(t_0)|
    [\hat{\rho}(rt),\hat{\rho}(r't')]|\Phi_\alpha(t_0)\rangle.
\end{align}
This formula is similar to the \textit{Kubo} formula used in calculation of linear response properties of a time-independent system, here the initial many-body sates $\ket{\Phi_{\alpha}(t_{0})}$ is defined at a particular time $t_{0}$.

\subsection{Fermi golden rule in Floquet systems}

To first order in the perturbing field
\begin{align}
    |\Psi_\mathrm{F}(t)\rangle 
    &=\left[ 1 -i\int_{t_0}^t dt_1 \hat{\Xi}_\mathrm{F}(t_1) \right] 
    |\Psi_\mathrm{F}(t_0)\rangle .
\end{align}
Returning to the Schr\"odinger picture, one finds
\begin{align}
    |\Psi(t)\rangle = \hat{U}_0(t,t_0)
    \left[ 1 -i\int_{t_0}^t dt_1 \hat{\Xi}_\mathrm{F}(t_1) \right] 
    |\Psi_\mathrm{F}(t_0)\rangle
\end{align}
With an initial state $|\Psi_\mathrm{F}(t_0)\rangle=|\Phi_\alpha(t_0)\rangle$, 
the transition amplitude
to a final state $|\Phi_\beta(t)\rangle$, with $\beta\neq\alpha$, is given by
\begin{align}
    \langle \Phi_\beta(t)|\Psi(t)\rangle 
    &=-i\int_{t_0}^t dt_1 \langle \Phi_\beta(t)|\hat{U}_0(t,t_1) \hat{\Xi}(t_1)
    |\Phi_\alpha(t_1)\rangle 
    \nonumber\\
    &\qquad\times e^{-iE_\alpha(t_1-t_0)}
    \nonumber\\
    &=-i\int_{t_0}^t dt_1 \langle \Phi_\beta(t_1)|\hat{\Xi}(t_1)
    |\Phi_\alpha(t_1)\rangle 
    \nonumber\\
    &\qquad\times e^{-iE_\beta(t-t_1)} e^{-iE_\alpha(t_1-t_0)},
\end{align}
where Eq. \eqref{eq:U0} has been utilized.
The transition probability is given by
\begin{align}
    &|\langle \Phi_\beta(t)|\Psi(t)\rangle|^2
    \nonumber\\
    &=\left|
    \int_{t_0}^t dt_1 \langle \Phi_\beta(t_1)|\hat{\Xi}(t_1)
    |\Phi_\alpha(t_1)\rangle e^{i(E_\beta-E_\alpha)t_1}
    \right|^2
\end{align}
For a harmonic perturbation with an arbitrary frequency $\omega$
\begin{align}
    \hat{\Xi}(t)=\hat{\Xi}e^{i\omega t}+\hat{\Xi}^\dagger e^{-i\omega t}.
\label{harmonic}
\end{align}
Let
\begin{align}
    a&=\int_{t_0}^t dt_1 \langle \Phi_\beta(t_1)|\hat{\Xi}
    |\Phi_\alpha(t_1)\rangle e^{i(\omega+\Delta_{\beta\alpha})t_1},
    \\
    b&=\int_{t_0}^t dt_1 \langle \Phi_\beta(t_1)|\hat{\Xi}^\dagger
    |\Phi_\alpha(t_1)\rangle e^{i(-\omega+\Delta_{\beta\alpha})t_1},
\end{align}
where $\Delta_{\beta\alpha}=E_\beta-E_\alpha$.
Since the Floquet modes are periodic in time, one may write the matrix
elements as Fourier series:
\begin{align}
    \langle \Phi_\beta(t_1)|\hat{\Xi}
    |\Phi_\alpha(t_1)\rangle &= \sum_n f^{\beta\alpha}_n e^{in\Omega t_1},
    \\
    \langle \Phi_\beta(t_1)|\hat{\Xi}^\dagger
    |\Phi_\alpha(t_1)\rangle &= \sum_n f^{\alpha\beta*}_n e^{-in\Omega t_1}.
\end{align}
One obtains
\begin{align}
    a&=\sum_n f_n^{\beta\alpha}
    \int_{t_0}^t dt_1  e^{i(\omega+n\Omega+\Delta_{\beta\alpha})t_1},
    \\
    b&=\sum_n f_n^{\alpha\beta*}\int_{t_0}^t dt_1 
    e^{i(-\omega+n\Omega+\Delta_{\beta\alpha})t_1}.
\end{align}
Assuming that $t_0=0$ one finds
\begin{align}
    a&=\sum_n f_n^{\beta\alpha} 
    \frac{e^{i(\omega+n\Omega+\Delta_{\beta\alpha})t}-1}
    {i(\omega+n\Omega+\Delta_{\beta\alpha})},
    \nonumber\\
    &=\sum_n f_n^{\beta\alpha} e^{i(\omega+n\Omega+\Delta_{\beta\alpha})t/2}
    \frac{2\sin{[(\omega+n\Omega+\Delta_{\beta\alpha})t/2]}}
    {\omega+n\Omega+\Delta_{\beta\alpha}},
    \\
    b&=\sum_n f_n^{\alpha\beta*}
    \frac{e^{i(-\omega+n\Omega+\Delta_{\beta\alpha})t}-1}
    {i(-\omega+n\Omega+\Delta_{\beta\alpha})}
    \nonumber\\
    &=\sum_n f_n^{\alpha\beta*}  e^{i(-\omega+n\Omega+\Delta_{\beta\alpha})t/2}
\frac{2\sin{[(\omega-n\Omega-\Delta_{\beta\alpha})t/2]}}
    {\omega-n\Omega-\Delta_{\beta\alpha}}.
\end{align}
The absolute value squared of the transition amplitude is
\begin{align}
    |\langle \Phi_\beta(t)|\Psi(t)\rangle|^2=|a+b|^2=|a|^2 +|b|^2 +a^*b+ab^*.
\end{align}
In the limit $t\rightarrow\infty$,
\begin{align}
    |a|^2&= 4\sum_n |f_n^{\beta\alpha}|^2 
    \frac{\sin^2{[(\omega+n\Omega+\Delta_{\beta\alpha})t/2]}}
    {(\omega+n\Omega+\Delta_{\beta\alpha})^2}
    \nonumber\\
    &=2\pi t \sum_n |f_n^{\beta\alpha}|^2
    \delta(\omega+n\Omega+\Delta_{\beta\alpha}),
\label{emission}
\end{align}
\begin{align}
    |b|^2 &= 4\sum_m |f_m^{\alpha\beta}|^2 
    \frac{\sin^2{[(\omega-m\Omega-\Delta_{\beta\alpha})t/2]}}
    {(\omega-m\Omega-\Delta_{\beta\alpha})^2},
    \nonumber\\
    &=2\pi t\sum_m |f_m^{\alpha\beta}|^2 
    \delta(\omega-m\Omega-\Delta_{\beta\alpha}).
\label{absorption}
\end{align}
If $2\Delta_{\beta\alpha}/\Omega=M$ (integer), then the cross terms
can contribute when $n=-m-2M$:
\begin{align}
    &a^*b+ab^*
    \nonumber\\
    &=4\pi t\sum_m  
    \text{Re}[f_m^{\alpha\beta}f_{-m-2M}^{\beta\alpha}]
    \delta(\omega-m\Omega-\Delta_{\beta\alpha}).
\label{cross}
\end{align}

Eqs. \eqref{emission}, \eqref{absorption}, and \eqref{cross} represent Umklapp in frequency space for absorption and emission processes associated to Floquet system. 

\subsection{Photoemission spectra in Floquet systems}

In the presence of an electromagnetic field, the electron momentum operator
$\mathbf{\hat{p}}$ is replaced by $\mathbf{\hat{p}}-(e/c)\mathbf{A}%
(\mathbf{\hat{r}},t)$\cite{Sakurai1993Modern} so that the kinetic energy becomes%
\begin{align}
&  \frac{1}{2m}[\mathbf{\hat{p}}-(e/c)\mathbf{A}(\mathbf{\hat{r}}%
,t)]^{2}\nonumber\\
&  =\frac{\mathbf{\hat{p}}^{2}}{2m}-\frac{e}{2mc}[\mathbf{\hat{p}\cdot
A}(\mathbf{\hat{r}},t)+\mathbf{A}(\mathbf{\hat{r}},t)\cdot\mathbf{\hat{p}%
]}\nonumber\\
&  \mathbf{+}\left(  \frac{e}{2mc}\right)  ^{2}\mathbf{A}^{2}(\mathbf{\hat{r}%
},t).
\end{align}
The last term is usually neglected since it is proportional to $1/c^{2}$ and
involves two-photon processes in first-order perturbation theory. It is
assumed that the vector potential represents a monochromatic field%

\begin{align}
\mathbf{A}  &  =2A_{0}\hat{\varepsilon}\cos\left(  \frac{\omega}%
{c}\mathbf{\hat{n}\cdot r}-\omega t\right) \nonumber\\
&  =A_{0}\hat{\varepsilon}\left[  e^{i\left(  \frac{\omega}{c}\mathbf{\hat
{n}\cdot r}-\omega t\right)  }+e^{-i\left(  \frac{\omega}{c}\mathbf{\hat
{n}\cdot r}-\omega t\right)  }\right]  .
\end{align}
$\hat{\varepsilon}$ is the light polarization and $\mathbf{\hat{n}}$ is the
propagation direction, and they are perpendicular to each other. Using the
Coulomb gauge $\nabla\cdot\mathbf{A}=0$ and treating%

\begin{equation}
-\frac{e}{2mc}\left(  \mathbf{\hat{p}\cdot\hat{A}+\hat{A}\cdot\hat{p}}\right)
=-\frac{e}{mc}\mathbf{\hat{A}\cdot\hat{p}}%
\end{equation}
as a perturbation representing the coupling between the photon and the
electron, the Fermi golden rule from the previous section can be directly
applied. One identifies%

\begin{equation}
\hat{V}=-\frac{eA_{0}}{mc}e^{i\frac{\omega}{c}\mathbf{\hat{n}\cdot\hat{x}}%
}\hat{\varepsilon}\cdot\mathbf{\hat{p}}.
\end{equation}
In the occupation number representation,%
\begin{equation}
\hat{V}(t)=\sum_{ij}(V_{ij}e^{i\omega t}+V_{ij}^{\ast}e^{-i\omega t})\hat
{c}_{i}^{\dag}\hat{c}_{j}.
\label{Vba}
\end{equation}

In a photoemission experiment, an electron is removed from an $N$-electron
system, which in the present case is a Floquet system. The calculation of the
photoemission spectrum is greatly simplified under the so-called sudden
approximation \cite{FADLEY1974225,Hedin_1999,Koralek_SA_2006} in which the photoelectron is assumed to be suddenly removed
from the system and completely decoupled from the remaining $(N-1)$-electron
system. This assumption is valid when the kinetic energy of the photoelectron
is large compared to its binding energy in the solid.

Let the initial state
be a Floquet mode $\left\vert \Phi_{\alpha}(t_0)\right\rangle $. Note that the
Greek alphabet is used to denote many-electron states whereas Roman letters
are used to denote one-particle base functions. 
Within the sudden approximation, the Floquet mode 
$\left\vert \Phi_{\beta}(t)\right\rangle $, 
which is coupled by the photon field to the initial state,
can be written as
\begin{equation}
\left\vert \Phi_{\beta}(t)\right\rangle =\hat{c}_{k}^{\dag}\left\vert
\Phi_{\gamma}^{N-1}(t)\right\rangle
\end{equation}
in which $\hat{c}_{k}^{\dag}$ represents the photoelectron with momentum
$k=(\mathbf{k},\sigma)$ and $\left\vert \Phi_{\gamma}^{N-1}(t)\right\rangle $
is an $(N-1)$-electron Floquet mode. It is evident that $\hat{c}_{i}^{\dag}$
in Eq. (\ref{Vba}) must be equal to $\hat{c}_{k}^{\dag}$ for otherwise the
matrix element would vanish. The transition matrix elements 
corresponding to an absorption process is then
\begin{align}
\left\langle \Phi_{\gamma}^{N-1}(t_1)\right\vert \sum_{j}V_{kj}^{\ast}\hat
{c}_{j}\left\vert \Phi_{\alpha}(t_1)\right\rangle
=\sum_m f_m^{k\alpha\gamma} e^{im\Omega t_1},
\end{align}
and the operator $\hat{\Xi}^\dagger$ in Eq. (\ref{harmonic}) can be
identified as
\begin{align}
    \hat{\Xi}^\dagger =\sum_{j}V_{kj}^{\ast}\hat{c}_{j}.
\end{align}
The Fermi golden rule for absorption in Eq. (\ref{absorption}) 
can now be applied to calculate the photocurrent
with energy $\epsilon_{k}$, the kinetic energy of the outgoing
photoelectron, yielding
\begin{align}
I(\epsilon_{k})  &  \varpropto
\sum_{\gamma m} |f_m^{k\alpha\gamma}|^2 
    \delta\left[\omega-\epsilon_k-m\Omega-(E^{N-1}_\gamma-E_\alpha)
    \right]. \label{eq:photocurrent}
\end{align}
The right-hand side of this equation, and thus also the photocurrent, is proportional to the imaginary part of the $l=0$ component of the Floquet Green function \cite{Uhrig2019}.
In general, both the initial and final states should be weighted by
taking into account the matrix element of the dipole operator between
these two states.
%
\begin{figure*}[h!]
\includegraphics[scale=0.87]{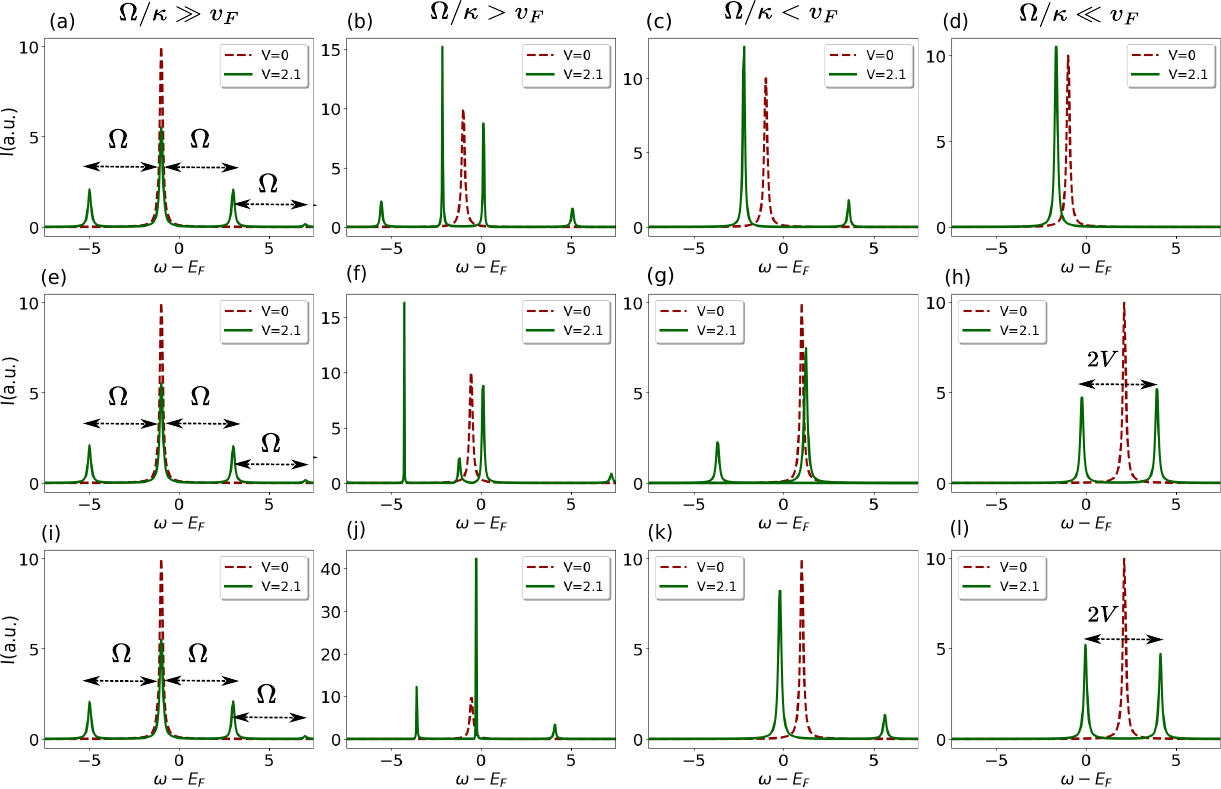}
	\caption{Spectral representation of photoelectric current of a noninteracting electron gas driven by an external periodic field  $V=V_{0}\text{cos}(\kappa x-\Omega t)$ (green) and nondriven (red). (a), (b), (c), and (d) at $\Gamma$ point, (e), (f), (g), and (h) at the left edge of the Brillouin zone, (i), (j), (k), and (l) at the right edge of the Brillouin zone.} 
	\label{fig:Photoelectric current}
\end{figure*}

\section{Photoelectric current of Periodically driven Electron Gas}
As an illustration of the formalism developed in this paper, we calculated the spectral behavior of the photoelectric current of a noninteracting electron gas driven by an external spatial-dependent time-periodic field $V(x,t)=V_{0}\text{cos}(\kappa x-\Omega t)$ as illustrated in Fig. \ref{fig:Response Floquet}. Here the equilibrium system has a parabolic band structure and both the spatial and temporal periodicity of the system are induced by the driving field. Since the initial system is periodic over both space and time with wave number $\kappa$ and frequency $\Omega$, we developed a combined Bloch-Floquet formalism to calculate the Green function and the subsequent photoelectric current. The dynamics of the initial system is characterized by two different velocities, the velocity of the driving field, which is in the form of a traveling wave, $v_{p}=\frac{\Omega}{\kappa}$, and the Fermi velocity $v_\mathrm{F}$ of the equilibrium system. The spectral nature of the current is shown in Fig. \ref{fig:Photoelectric current} for different values of the ratio $v_{p} /v_\mathrm{F}$ and at different $k$ points in the Brillouin zone. For simplicity we choose $m_{e}=1$ and $\hbar=1$. All parameters, $V_{0}$, $\kappa$, and $\Omega$, and the $\omega$ axis are scaled in the unit of the Fermi energy $E_\mathrm{F}$ of the equilibrium system. For all the nonequilibrium calculations, we set the strength of the driving field $V_{0}=2.1E_\mathrm{F}$.
\subsection{High frequency limit $\Omega /\kappa \gg v_\mathrm{F}$}
In this limit the system is governed by the Floquet theory and the Bloch part becomes negligible. This case is shown for $\Omega=40\kappa$ and $v_\mathrm{F} = \sqrt{2}$ in Fig.~\ref{fig:Photoelectric current}(a-e-i). In equilibrium ($V=0$), the photocurrent has only one peak centered at the non-interacting energy, which corresponds to the non-interacting band, $E(k) \approx -1$. Since the size of the Brillouin zone is very small, the profile remains almost the same irrespective of which k point in the Brillouin zone we are looking at. In presence of the driving field, spectral weight is transferred to side peaks shifted by integer multiples of the Floquet frequency $\Omega$. 
\subsection{Low frequency limit $\Omega /\kappa \ll v_\mathrm{F}$}
In the low-frequency limit, the characteristic velocity of the driving field $v_{p}$ becomes very small so the system returns to a usual time-independent Bloch system and the Floquet part becomes negligible. This case is shown in Fig.~\ref{fig:Photoelectric current}(d,h,l), for $\kappa=25\Omega$ and $v_\mathrm{F}=\sqrt{2}$. At the $\Gamma$ point both the equilibrium and nonequilibrium current profiles have one peak. At the edges of the Brillouin zone the current profile of the driven system has two peaks separated by energy $\sim 2V_{0}$, the usual band gand appearing due to a periodic potential. The peak heights and the positions are not exactly symmetric about the equilibrium peak due to the nonzero value of $\Omega$. 
\subsection{Intermediate regime}
The intermediate regime combines the Floquet and Bloch pictures, as illustrated by the examples $\Omega = 2\kappa$ shown in Fig.~\ref{fig:Photoelectric current}(b-f-j) and $\Omega=\kappa$ in Fig.~\ref{fig:Photoelectric current}(c-g-k). A crossover from the previously discussed Floquet to Bloch physics is visible. The peaks all have different heights and at the edges of the Brillouin zone their gap is not equal to $2V_{0}$. Furthermore, in these two cases the current profile is not symmetric about the $\Gamma$ point, i.e., panels (f) and (j) are different, since the driving is a traveling wave which breaks both time-reversal and spatial inversion symmetry and therefore breaks the Kramers degeneracy. 
%
\section{Conclusion}

We have developed a linear response formalism for Floquet systems
analogous to the Kubo formalism for equilibrium systems.
As an application, we have derived the formula for photoemission spectra within the sudden approximation and shown that it is proportional to the $l=0$ component of the imaginary part of the
Green function. This is consistent with an earlier finding that the $l=0$ component is positive definite. The formula provides a well-defined means for making a comparison between theory and experiment. We have considered a noninteracting homogeneous
electron gas to illustrate our formalism. With that, we analyzed the parameter dependence of the photoemission spectra of that controllable Floquet system, which is the signature of a tunable band structure. This can have potential application in Floquet engineering.

\section{Acknowledgment}
The authors thank Mikhail Katsnelson and Chin Shen Ong for useful discussions.
The authors acknowledge the Swedish Research Council (Vetenskapsprådet, VR)
Grant number 2021-04498\_3
and the Crafoord Foundation (grant number 20220901) for financial support. 

\appendix

\section{Homogeneous Electron gas under time-periodic potential}
As an illustration, we consider a noninteracting homogeneous electron gas in the presence of a time-periodic but spatially homogeneous potential. A spatially homogeneous potential provides substantial simplifications since it does not mix the different plane wave states of the electron gas. Thus, the Green function can be calculated analytically. 
Let the periodic potential be $V(t)=V_0\cos(\Omega t)$.
Then the Floquet modes fulfill
the following eigenvalue problem:
\begin{equation}
\left[  -\frac{1}{2}\nabla^{2}+V_0\cos(\Omega t)-i\partial_{t}\right]
\phi_{\mathbf{k}\alpha}(rt)=\varepsilon_{\mathbf{k}\alpha}\phi_{\mathbf{k}%
\alpha}(rt).
\end{equation}
The spatial part of $\phi_{\mathbf{k}\alpha}(rt)$ must be a plane wave so that
\begin{equation}
\phi_{\mathbf{k}\alpha}(rt)=\frac{\exp(i\mathbf{k}\cdot\mathbf{r})}
{\sqrt{L^3}}\phi_{\alpha}(t),
\end{equation}
where $L^3$ is the space volume and
$\phi_{\alpha}(t)$ is periodic with period $T$:
\begin{equation}
\phi_{\alpha}(t+T)=\phi_{\alpha}(t).
\end{equation}
One then finds
\begin{equation}
\left[  V_0\cos(\Omega t)-i\partial_{t}\right]  \phi_{\alpha}(t)=\widetilde
{\varepsilon}_{\alpha}\phi_{\alpha}(t)
\end{equation}
in which
\begin{equation}
\widetilde{\varepsilon}_{\alpha}=\varepsilon_{\mathbf{k}\alpha}-\frac{1}
{2}k^{2}.
\end{equation}
The solution is given by
\begin{align}
\phi_{\alpha}(t)  &  =\phi_{\alpha}(0)\exp\left\{  -i\int_{0}^{t}dt^{\prime
}\left[  V_0\cos(\Omega t^{\prime})-\widetilde{\varepsilon}_{\alpha}\right]
\right\} \nonumber\\
&  =\phi_{\alpha}(0)\exp\left\{  -i\left[  \frac{V_0}{\Omega}\sin(\Omega
t)-\widetilde{\varepsilon}_{\alpha}t\right]  \right\}  .
\end{align}
The constant $\phi_{n}(0)$ is equal to unity. The eigenvalues are determined
by requiring that $\phi_{\alpha}(t+T)=\phi_{\alpha}(t)$, i.e.,
\begin{equation}
\widetilde{\varepsilon}_{\alpha}T=2\pi\times\text{integer}\rightarrow
\widetilde{\varepsilon}_{\alpha}=\alpha\Omega\rightarrow\varepsilon
_{\mathbf{k}\alpha}=\frac{1}{2}k^{2}+\alpha\Omega,
\end{equation}
where $\alpha=0,\pm1,\pm2,\cdots$. The Floquet modes become
\begin{equation}
\phi_{\mathbf{k}\alpha}(rt)=\frac{\exp(i\mathbf{k}\cdot\mathbf{r)}}
{\sqrt{L^3}}\exp\left\{  -i\left[  \frac{V_0}{\Omega}\sin(\Omega
t)-\alpha\Omega t\right]  \right\}
\end{equation}
and the Floquet states are
\begin{align}
\psi_{\mathbf{k}\alpha}(rt)  &  =\exp(-i\varepsilon_{\mathbf{k}\alpha}%
t)\phi_{\mathbf{k}\alpha}(rt)\nonumber\\
&  =\frac{\exp(i\mathbf{k}\cdot\mathbf{r)}}{\sqrt{L^3}}\exp\left\{
-i\left[  \frac{V_0}{\Omega}\sin(\Omega t)+\frac{1}{2}k^{2}t\right]  \right\}  ,
\end{align}
In this simple case, the Floquet states are independent of $\alpha$.

The retarded Green function is then given by%
\begin{equation}
G^{\text{R}}(rt,r^{\prime}t^{\prime})=-i\theta(t-t^{\prime})\sum_{\mathbf{k}%
}\psi_{\mathbf{k}\alpha}(rt)\psi_{\mathbf{k}\alpha}^{\ast}(r^{\prime}%
t^{\prime})
\end{equation}
which fulfills the equation of motion,
\begin{equation}
\left[  i\partial_{t}+\frac{1}{2}\nabla^{2}-V_0\cos(\Omega t)\right]
G^{\text{R}}(rt,r^{\prime}t^{\prime})=\delta(r-r^{\prime})\delta(t-t^{\prime
}).
\end{equation}


\bibliographystyle{apsrev4-1.bst}
\bibliography{cite.bib}

\begin{thebibliography}{22}%
\makeatletter
\providecommand \@ifxundefined [1]{%
 \@ifx{#1\undefined}
}%
\providecommand \@ifnum [1]{%
 \ifnum #1\expandafter \@firstoftwo
 \else \expandafter \@secondoftwo
 \fi
}%
\providecommand \@ifx [1]{%
 \ifx #1\expandafter \@firstoftwo
 \else \expandafter \@secondoftwo
 \fi
}%
\providecommand \natexlab [1]{#1}%
\providecommand \enquote  [1]{``#1''}%
\providecommand \bibnamefont  [1]{#1}%
\providecommand \bibfnamefont [1]{#1}%
\providecommand \citenamefont [1]{#1}%
\providecommand \href@noop [0]{\@secondoftwo}%
\providecommand \href [0]{\begingroup \@sanitize@url \@href}%
\providecommand \@href[1]{\@@startlink{#1}\@@href}%
\providecommand \@@href[1]{\endgroup#1\@@endlink}%
\providecommand \@sanitize@url [0]{\catcode `\\12\catcode `\$12\catcode
  `\&12\catcode `\#12\catcode `\^12\catcode `\_12\catcode `\%12\relax}%
\providecommand \@@startlink[1]{}%
\providecommand \@@endlink[0]{}%
\providecommand \url  [0]{\begingroup\@sanitize@url \@url }%
\providecommand \@url [1]{\endgroup\@href {#1}{\urlprefix }}%
\providecommand \urlprefix  [0]{URL }%
\providecommand \Eprint [0]{\href }%
\providecommand \doibase [0]{http://dx.doi.org/}%
\providecommand \selectlanguage [0]{\@gobble}%
\providecommand \bibinfo  [0]{\@secondoftwo}%
\providecommand \bibfield  [0]{\@secondoftwo}%
\providecommand \translation [1]{[#1]}%
\providecommand \BibitemOpen [0]{}%
\providecommand \bibitemStop [0]{}%
\providecommand \bibitemNoStop [0]{.\EOS\space}%
\providecommand \EOS [0]{\spacefactor3000\relax}%
\providecommand \BibitemShut  [1]{\csname bibitem#1\endcsname}%
\let\auto@bib@innerbib\@empty
\bibitem [{\citenamefont {Broers}\ and\ \citenamefont
  {Mathey}(2022)}]{Broers2022Graphene}%
  \BibitemOpen
  \bibfield  {author} {\bibinfo {author} {\bibfnamefont {L.}~\bibnamefont
  {Broers}}\ and\ \bibinfo {author} {\bibfnamefont {L.}~\bibnamefont
  {Mathey}},\ }\href {\doibase 10.1103/PhysRevResearch.4.013057} {\bibfield
  {journal} {\bibinfo  {journal} {Phys. Rev. Res.}\ }\textbf {\bibinfo {volume}
  {4}},\ \bibinfo {pages} {013057} (\bibinfo {year} {2022})}\BibitemShut
  {NoStop}%
\bibitem [{\citenamefont {Broers}\ and\ \citenamefont
  {Mathey}(2021)}]{Broers2021Graphene}%
  \BibitemOpen
  \bibfield  {author} {\bibinfo {author} {\bibfnamefont {L.}~\bibnamefont
  {Broers}}\ and\ \bibinfo {author} {\bibfnamefont {L.}~\bibnamefont
  {Mathey}},\ }\href {\doibase 10.1038/s42005-021-00746-6} {\bibfield
  {journal} {\bibinfo  {journal} {Communications Physics}\ }\textbf {\bibinfo
  {volume} {4}} (\bibinfo {year} {2021}),\
  10.1038/s42005-021-00746-6}\BibitemShut {NoStop}%
\bibitem [{\citenamefont {Grifoni}\ and\ \citenamefont
  {Hänggi}(1998)}]{GRIFONI1998229}%
  \BibitemOpen
  \bibfield  {author} {\bibinfo {author} {\bibfnamefont {M.}~\bibnamefont
  {Grifoni}}\ and\ \bibinfo {author} {\bibfnamefont {P.}~\bibnamefont
  {Hänggi}},\ }\href {\doibase https://doi.org/10.1016/S0370-1573(98)00022-2}
  {\bibfield  {journal} {\bibinfo  {journal} {Physics Reports}\ }\textbf
  {\bibinfo {volume} {304}},\ \bibinfo {pages} {229} (\bibinfo {year}
  {1998})}\BibitemShut {NoStop}%
\bibitem [{\citenamefont {Aoki}\ \emph {et~al.}(2014)\citenamefont {Aoki},
  \citenamefont {Tsuji}, \citenamefont {Eckstein}, \citenamefont {Kollar},
  \citenamefont {Oka},\ and\ \citenamefont {Werner}}]{RevModPhys.86.779}%
  \BibitemOpen
  \bibfield  {author} {\bibinfo {author} {\bibfnamefont {H.}~\bibnamefont
  {Aoki}}, \bibinfo {author} {\bibfnamefont {N.}~\bibnamefont {Tsuji}},
  \bibinfo {author} {\bibfnamefont {M.}~\bibnamefont {Eckstein}}, \bibinfo
  {author} {\bibfnamefont {M.}~\bibnamefont {Kollar}}, \bibinfo {author}
  {\bibfnamefont {T.}~\bibnamefont {Oka}}, \ and\ \bibinfo {author}
  {\bibfnamefont {P.}~\bibnamefont {Werner}},\ }\href {\doibase
  10.1103/RevModPhys.86.779} {\bibfield  {journal} {\bibinfo  {journal} {Rev.
  Mod. Phys.}\ }\textbf {\bibinfo {volume} {86}},\ \bibinfo {pages} {779}
  (\bibinfo {year} {2014})}\BibitemShut {NoStop}%
\bibitem [{\citenamefont {Oka}\ and\ \citenamefont
  {Aoki}(2009)}]{Oka2009Hall_Effect}%
  \BibitemOpen
  \bibfield  {author} {\bibinfo {author} {\bibfnamefont {T.}~\bibnamefont
  {Oka}}\ and\ \bibinfo {author} {\bibfnamefont {H.}~\bibnamefont {Aoki}},\
  }\href {\doibase 10.1103/PhysRevB.79.081406} {\bibfield  {journal} {\bibinfo
  {journal} {Phys. Rev. B}\ }\textbf {\bibinfo {volume} {79}},\ \bibinfo
  {pages} {081406} (\bibinfo {year} {2009})}\BibitemShut {NoStop}%
\bibitem [{\citenamefont {Fausti}\ \emph {et~al.}(2011)\citenamefont {Fausti},
  \citenamefont {Tobey}, \citenamefont {Dean}, \citenamefont {Kaiser},
  \citenamefont {Dienst}, \citenamefont {Hoffmann}, \citenamefont {Pyon},
  \citenamefont {Takayama}, \citenamefont {Takagi},\ and\ \citenamefont
  {Cavalleri}}]{fausti2011light}%
  \BibitemOpen
  \bibfield  {author} {\bibinfo {author} {\bibfnamefont {D.}~\bibnamefont
  {Fausti}}, \bibinfo {author} {\bibfnamefont {R.}~\bibnamefont {Tobey}},
  \bibinfo {author} {\bibfnamefont {N.}~\bibnamefont {Dean}}, \bibinfo {author}
  {\bibfnamefont {S.}~\bibnamefont {Kaiser}}, \bibinfo {author} {\bibfnamefont
  {A.}~\bibnamefont {Dienst}}, \bibinfo {author} {\bibfnamefont {M.~C.}\
  \bibnamefont {Hoffmann}}, \bibinfo {author} {\bibfnamefont {S.}~\bibnamefont
  {Pyon}}, \bibinfo {author} {\bibfnamefont {T.}~\bibnamefont {Takayama}},
  \bibinfo {author} {\bibfnamefont {H.}~\bibnamefont {Takagi}}, \ and\ \bibinfo
  {author} {\bibfnamefont {A.}~\bibnamefont {Cavalleri}},\ }\href@noop {}
  {\bibfield  {journal} {\bibinfo  {journal} {Science}\ }\textbf {\bibinfo
  {volume} {331}},\ \bibinfo {pages} {189} (\bibinfo {year}
  {2011})}\BibitemShut {NoStop}%
\bibitem [{\citenamefont {Bloch}\ \emph {et~al.}(2022)\citenamefont {Bloch},
  \citenamefont {Cavalleri}, \citenamefont {Galitski}, \citenamefont {Hafezi},\
  and\ \citenamefont {Rubio}}]{Bloch_2022}%
  \BibitemOpen
  \bibfield  {author} {\bibinfo {author} {\bibfnamefont {J.}~\bibnamefont
  {Bloch}}, \bibinfo {author} {\bibfnamefont {A.}~\bibnamefont {Cavalleri}},
  \bibinfo {author} {\bibfnamefont {V.}~\bibnamefont {Galitski}}, \bibinfo
  {author} {\bibfnamefont {M.}~\bibnamefont {Hafezi}}, \ and\ \bibinfo {author}
  {\bibfnamefont {A.}~\bibnamefont {Rubio}},\ }\href {\doibase
  10.1038/s41586-022-04726-w} {\bibfield  {journal} {\bibinfo  {journal}
  {Nature}\ }\textbf {\bibinfo {volume} {606}},\ \bibinfo {pages} {41–48}
  (\bibinfo {year} {2022})}\BibitemShut {NoStop}%
\bibitem [{\citenamefont {Eckstein}(2021)}]{ECKSTEIN2021147108}%
  \BibitemOpen
  \bibfield  {author} {\bibinfo {author} {\bibfnamefont {M.}~\bibnamefont
  {Eckstein}},\ }\href {\doibase https://doi.org/10.1016/j.elspec.2021.147108}
  {\bibfield  {journal} {\bibinfo  {journal} {Journal of Electron Spectroscopy
  and Related Phenomena}\ }\textbf {\bibinfo {volume} {253}},\ \bibinfo {pages}
  {147108} (\bibinfo {year} {2021})}\BibitemShut {NoStop}%
\bibitem [{\citenamefont {Beaurepaire}\ \emph {et~al.}(1996)\citenamefont
  {Beaurepaire}, \citenamefont {Merle}, \citenamefont {Daunois},\ and\
  \citenamefont {Bigot}}]{PhysRevLett.76.4250}%
  \BibitemOpen
  \bibfield  {author} {\bibinfo {author} {\bibfnamefont {E.}~\bibnamefont
  {Beaurepaire}}, \bibinfo {author} {\bibfnamefont {J.-C.}\ \bibnamefont
  {Merle}}, \bibinfo {author} {\bibfnamefont {A.}~\bibnamefont {Daunois}}, \
  and\ \bibinfo {author} {\bibfnamefont {J.-Y.}\ \bibnamefont {Bigot}},\ }\href
  {\doibase 10.1103/PhysRevLett.76.4250} {\bibfield  {journal} {\bibinfo
  {journal} {Phys. Rev. Lett.}\ }\textbf {\bibinfo {volume} {76}},\ \bibinfo
  {pages} {4250} (\bibinfo {year} {1996})}\BibitemShut {NoStop}%
\bibitem [{\citenamefont {Kirilyuk}\ \emph {et~al.}(2010)\citenamefont
  {Kirilyuk}, \citenamefont {Kimel},\ and\ \citenamefont
  {Rasing}}]{RevModPhys.82.2731}%
  \BibitemOpen
  \bibfield  {author} {\bibinfo {author} {\bibfnamefont {A.}~\bibnamefont
  {Kirilyuk}}, \bibinfo {author} {\bibfnamefont {A.~V.}\ \bibnamefont {Kimel}},
  \ and\ \bibinfo {author} {\bibfnamefont {T.}~\bibnamefont {Rasing}},\ }\href
  {\doibase 10.1103/RevModPhys.82.2731} {\bibfield  {journal} {\bibinfo
  {journal} {Rev. Mod. Phys.}\ }\textbf {\bibinfo {volume} {82}},\ \bibinfo
  {pages} {2731} (\bibinfo {year} {2010})}\BibitemShut {NoStop}%
\bibitem [{\citenamefont {Trevisan}\ \emph {et~al.}(2022)\citenamefont
  {Trevisan}, \citenamefont {Arribi}, \citenamefont {Heinonen}, \citenamefont
  {Slager},\ and\ \citenamefont {Orth}}]{Trevisan2022PRL}%
  \BibitemOpen
  \bibfield  {author} {\bibinfo {author} {\bibfnamefont {T.~V.}\ \bibnamefont
  {Trevisan}}, \bibinfo {author} {\bibfnamefont {P.~V.}\ \bibnamefont
  {Arribi}}, \bibinfo {author} {\bibfnamefont {O.}~\bibnamefont {Heinonen}},
  \bibinfo {author} {\bibfnamefont {R.-J.}\ \bibnamefont {Slager}}, \ and\
  \bibinfo {author} {\bibfnamefont {P.~P.}\ \bibnamefont {Orth}},\ }\href
  {\doibase 10.1103/PhysRevLett.128.066602} {\bibfield  {journal} {\bibinfo
  {journal} {Phys. Rev. Lett.}\ }\textbf {\bibinfo {volume} {128}},\ \bibinfo
  {pages} {066602} (\bibinfo {year} {2022})}\BibitemShut {NoStop}%
\bibitem [{\citenamefont {Rechtsman}\ \emph {et~al.}(2013)\citenamefont
  {Rechtsman}, \citenamefont {Zeuner}, \citenamefont {Plotnik}, \citenamefont
  {Lumer}, \citenamefont {Podolsky}, \citenamefont {Dreisow}, \citenamefont
  {Nolte}, \citenamefont {Segev},\ and\ \citenamefont
  {Szameit}}]{rechtsman2013photonic}%
  \BibitemOpen
  \bibfield  {author} {\bibinfo {author} {\bibfnamefont {M.~C.}\ \bibnamefont
  {Rechtsman}}, \bibinfo {author} {\bibfnamefont {J.~M.}\ \bibnamefont
  {Zeuner}}, \bibinfo {author} {\bibfnamefont {Y.}~\bibnamefont {Plotnik}},
  \bibinfo {author} {\bibfnamefont {Y.}~\bibnamefont {Lumer}}, \bibinfo
  {author} {\bibfnamefont {D.}~\bibnamefont {Podolsky}}, \bibinfo {author}
  {\bibfnamefont {F.}~\bibnamefont {Dreisow}}, \bibinfo {author} {\bibfnamefont
  {S.}~\bibnamefont {Nolte}}, \bibinfo {author} {\bibfnamefont
  {M.}~\bibnamefont {Segev}}, \ and\ \bibinfo {author} {\bibfnamefont
  {A.}~\bibnamefont {Szameit}},\ }\href@noop {} {\bibfield  {journal} {\bibinfo
   {journal} {Nature}\ }\textbf {\bibinfo {volume} {496}},\ \bibinfo {pages}
  {196} (\bibinfo {year} {2013})}\BibitemShut {NoStop}%
\bibitem [{\citenamefont {Grushin}\ \emph {et~al.}(2014)\citenamefont
  {Grushin}, \citenamefont {G\'omez-Le\'on},\ and\ \citenamefont
  {Neupert}}]{Grushin2014CI}%
  \BibitemOpen
  \bibfield  {author} {\bibinfo {author} {\bibfnamefont {A.~G.}\ \bibnamefont
  {Grushin}}, \bibinfo {author} {\bibfnamefont {A.}~\bibnamefont
  {G\'omez-Le\'on}}, \ and\ \bibinfo {author} {\bibfnamefont {T.}~\bibnamefont
  {Neupert}},\ }\href {\doibase 10.1103/PhysRevLett.112.156801} {\bibfield
  {journal} {\bibinfo  {journal} {Phys. Rev. Lett.}\ }\textbf {\bibinfo
  {volume} {112}},\ \bibinfo {pages} {156801} (\bibinfo {year}
  {2014})}\BibitemShut {NoStop}%
\bibitem [{\citenamefont {Yates}\ \emph {et~al.}(2016)\citenamefont {Yates},
  \citenamefont {Lemonik},\ and\ \citenamefont {Mitra}}]{Yates2016CI}%
  \BibitemOpen
  \bibfield  {author} {\bibinfo {author} {\bibfnamefont {D.~J.}\ \bibnamefont
  {Yates}}, \bibinfo {author} {\bibfnamefont {Y.}~\bibnamefont {Lemonik}}, \
  and\ \bibinfo {author} {\bibfnamefont {A.}~\bibnamefont {Mitra}},\ }\href
  {\doibase 10.1103/PhysRevB.94.205422} {\bibfield  {journal} {\bibinfo
  {journal} {Phys. Rev. B}\ }\textbf {\bibinfo {volume} {94}},\ \bibinfo
  {pages} {205422} (\bibinfo {year} {2016})}\BibitemShut {NoStop}%
\bibitem [{\citenamefont {Li}\ \emph {et~al.}(2018)\citenamefont {Li},
  \citenamefont {Liu},\ and\ \citenamefont {Yao}}]{Li_2018CI}%
  \BibitemOpen
  \bibfield  {author} {\bibinfo {author} {\bibfnamefont {S.}~\bibnamefont
  {Li}}, \bibinfo {author} {\bibfnamefont {C.-C.}\ \bibnamefont {Liu}}, \ and\
  \bibinfo {author} {\bibfnamefont {Y.}~\bibnamefont {Yao}},\ }\href {\doibase
  10.1088/1367-2630/aab2c7} {\bibfield  {journal} {\bibinfo  {journal} {New
  Journal of Physics}\ }\textbf {\bibinfo {volume} {20}},\ \bibinfo {pages}
  {033025} (\bibinfo {year} {2018})}\BibitemShut {NoStop}%
\bibitem [{\citenamefont {Uhrig}\ \emph {et~al.}(2019)\citenamefont {Uhrig},
  \citenamefont {Kalthoff},\ and\ \citenamefont {Freericks}}]{Uhrig2019}%
  \BibitemOpen
  \bibfield  {author} {\bibinfo {author} {\bibfnamefont {G.~S.}\ \bibnamefont
  {Uhrig}}, \bibinfo {author} {\bibfnamefont {M.~H.}\ \bibnamefont {Kalthoff}},
  \ and\ \bibinfo {author} {\bibfnamefont {J.~K.}\ \bibnamefont {Freericks}},\
  }\href {\doibase 10.1103/PhysRevLett.122.130604} {\bibfield  {journal}
  {\bibinfo  {journal} {Phys. Rev. Lett.}\ }\textbf {\bibinfo {volume} {122}},\
  \bibinfo {pages} {130604} (\bibinfo {year} {2019})}\BibitemShut {NoStop}%
\bibitem [{\citenamefont {Stefanucci}\ and\ \citenamefont {van
  Leeuwen}(2010)}]{Stefanucci}%
  \BibitemOpen
  \bibfield  {author} {\bibinfo {author} {\bibfnamefont {G.}~\bibnamefont
  {Stefanucci}}\ and\ \bibinfo {author} {\bibfnamefont {R.}~\bibnamefont {van
  Leeuwen}},\ }\href {\doibase 10.1017/CBO9781139023979} {\emph {\bibinfo
  {title} {Nonequilibrium Many-Body Theory of Quantum Systems: A Modern
  Introduction}}}\ (\bibinfo {year} {2010})\ pp.\ \bibinfo {pages}
  {1--600}\BibitemShut {NoStop}%
\bibitem [{\citenamefont {Tsuji}\ \emph {et~al.}(2008)\citenamefont {Tsuji},
  \citenamefont {Oka},\ and\ \citenamefont {Aoki}}]{Tsuji_2008}%
  \BibitemOpen
  \bibfield  {author} {\bibinfo {author} {\bibfnamefont {N.}~\bibnamefont
  {Tsuji}}, \bibinfo {author} {\bibfnamefont {T.}~\bibnamefont {Oka}}, \ and\
  \bibinfo {author} {\bibfnamefont {H.}~\bibnamefont {Aoki}},\ }\href {\doibase
  10.1103/PhysRevB.78.235124} {\bibfield  {journal} {\bibinfo  {journal} {Phys.
  Rev. B}\ }\textbf {\bibinfo {volume} {78}},\ \bibinfo {pages} {235124}
  (\bibinfo {year} {2008})}\BibitemShut {NoStop}%
\bibitem [{\citenamefont {Sakurai}(1993)}]{Sakurai1993Modern}%
  \BibitemOpen
  \bibfield  {author} {\bibinfo {author} {\bibfnamefont {J.~J.}\ \bibnamefont
  {Sakurai}},\ }\href {http://www.worldcat.org/isbn/0201539292} {\emph
  {\bibinfo {title} {Modern Quantum Mechanics (Revised Edition)}}},\ \bibinfo
  {edition} {1st}\ ed.\ (\bibinfo  {publisher} {Addison Wesley},\ \bibinfo
  {year} {1993})\BibitemShut {NoStop}%
\bibitem [{\citenamefont {Fadley}(1974)}]{FADLEY1974225}%
  \BibitemOpen
  \bibfield  {author} {\bibinfo {author} {\bibfnamefont {C.}~\bibnamefont
  {Fadley}},\ }\href {\doibase https://doi.org/10.1016/0009-2614(74)89123-2}
  {\bibfield  {journal} {\bibinfo  {journal} {Chemical Physics Letters}\
  }\textbf {\bibinfo {volume} {25}},\ \bibinfo {pages} {225} (\bibinfo {year}
  {1974})}\BibitemShut {NoStop}%
\bibitem [{\citenamefont {Hedin}(1999)}]{Hedin_1999}%
  \BibitemOpen
  \bibfield  {author} {\bibinfo {author} {\bibfnamefont {L.}~\bibnamefont
  {Hedin}},\ }\href {\doibase 10.1088/0953-8984/11/42/201} {\bibfield
  {journal} {\bibinfo  {journal} {Journal of Physics: Condensed Matter}\
  }\textbf {\bibinfo {volume} {11}},\ \bibinfo {pages} {R489} (\bibinfo {year}
  {1999})}\BibitemShut {NoStop}%
\bibitem [{\citenamefont {Koralek}\ \emph {et~al.}(2006)\citenamefont
  {Koralek}, \citenamefont {Douglas}, \citenamefont {Plumb}, \citenamefont
  {Sun}, \citenamefont {Fedorov}, \citenamefont {Murnane}, \citenamefont
  {Kapteyn}, \citenamefont {Cundiff}, \citenamefont {Aiura}, \citenamefont
  {Oka}, \citenamefont {Eisaki},\ and\ \citenamefont
  {Dessau}}]{Koralek_SA_2006}%
  \BibitemOpen
  \bibfield  {author} {\bibinfo {author} {\bibfnamefont {J.~D.}\ \bibnamefont
  {Koralek}}, \bibinfo {author} {\bibfnamefont {J.~F.}\ \bibnamefont
  {Douglas}}, \bibinfo {author} {\bibfnamefont {N.~C.}\ \bibnamefont {Plumb}},
  \bibinfo {author} {\bibfnamefont {Z.}~\bibnamefont {Sun}}, \bibinfo {author}
  {\bibfnamefont {A.~V.}\ \bibnamefont {Fedorov}}, \bibinfo {author}
  {\bibfnamefont {M.~M.}\ \bibnamefont {Murnane}}, \bibinfo {author}
  {\bibfnamefont {H.~C.}\ \bibnamefont {Kapteyn}}, \bibinfo {author}
  {\bibfnamefont {S.~T.}\ \bibnamefont {Cundiff}}, \bibinfo {author}
  {\bibfnamefont {Y.}~\bibnamefont {Aiura}}, \bibinfo {author} {\bibfnamefont
  {K.}~\bibnamefont {Oka}}, \bibinfo {author} {\bibfnamefont {H.}~\bibnamefont
  {Eisaki}}, \ and\ \bibinfo {author} {\bibfnamefont {D.~S.}\ \bibnamefont
  {Dessau}},\ }\href {\doibase 10.1103/PhysRevLett.96.017005} {\bibfield
  {journal} {\bibinfo  {journal} {Phys. Rev. Lett.}\ }\textbf {\bibinfo
  {volume} {96}},\ \bibinfo {pages} {017005} (\bibinfo {year}
  {2006})}\BibitemShut {NoStop}%
\end{thebibliography}%


\begin{thebibliography}{9}

\bibitem {}F. Aryasetiawan and O. Gunnarsson, Rep. Prog. Phys. \textbf{61},
237 (1998).

\bibitem {georges1996}A. Georges, G. Kotliar, W. Krauth, and M. J. Rozenberg,
Rev. Mod. Phys. \textbf{68}, 13 (1996).

\bibitem {hedin1965}L. Hedin, Phys. Rev. \textbf{139}, A796 (1965).

\bibitem {grifoni1998}M. Grifoni and P. H\"{a}nggi, Physics Reports
\textbf{304}, 229-354 (1998).

\bibitem {negele}J. W. Negele and H. Orland, \emph{Quantum Many-Particle
Systems}, (Westview Press, Boulder, Colorado, 1998).

\bibitem {sakurai}J. J. Sakurai, \emph{Modern Quantum Mechanics, }Revised
edition\emph{ }(Addison Wesley, 1994).

\bibitem {stefanucci}G. Stefanucci and R. van Leeuwen, \emph{Nonequilibrium
Many-Body Theory of Quantum Systems}, (Cambridge University Press, Cambridge 2013).

\bibitem {tsuji2008}N. Tsuji, T. Oka, and H. Aoki, Phys. Rev. B \textbf{78},
235124 (2008).

\bibitem {uhrig2019}G. S. Uhrig, M. H. Kalthoff, and J. K. Freericks, Phys.
Rev. Lett. \textbf{122}, 130604 (2019).

\bibitem{Beaurepaire96}E. Beaurepaire, J.C. Merle, A. Daunois, and J.Y. Bigot, Phys. Rev. Lett \textbf{76}, 22 (1996) 

\bibitem{Kirilyuk2010}A. Kirilyuk, A. V. Kimel, and T. Rasing, Rev. Mod. Phys. \textbf{82}, 2731 (2010)

\bibitem{Bloch2022}Bloch, J., Cavalleri, A., Galitski, V. et al. \textit{Strongly correlated electron–photon systems.} Nature \textbf{606}, 41–48 (2022)

\bibitem{Eckstein2021}M. Eckstein, Journal of Electron Spectroscopy and Related Phenomena
 \textbf{253}, 147108 (2021)
\end{thebibliography}

\end{document}